\documentclass{elsart}
\usepackage{natbib}
\usepackage{graphics}
\newcommand{\bc}{\begin{center}}
\newcommand{\ec}{\end{center}}
\newcommand{\be}{\begin{equation}}
\newcommand{\ee}{\end{equation}}
\newcommand{\ba}{\begin{array}}
\newcommand{\ea}{\end{array}}
\newcommand{\bea}{\begin{eqnarray}}
\newcommand{\eea}{\end{eqnarray}}
\def\ga{\mathrel{\mathchoice {\vcenter{\offinterlineskip\halign{\hfil
$\displaystyle##$\hfil\cr>\cr\sim\cr}}}
{\vcenter{\offinterlineskip\halign{\hfil$\textstyle##$\hfil\cr
>\cr\sim\cr}}}
{\vcenter{\offinterlineskip\halign{\hfil$\scriptstyle##$\hfil\cr
>\cr\sim\cr}}}
{\vcenter{\offinterlineskip\halign{\hfil$\scriptscriptstyle##$\hfil\cr
>\cr\sim\cr}}}}}

\begin{document}
\runauthor{Voshchinnikov et al.}
\begin{frontmatter}
\title{Extinction and polarization
of radiation by absorbing spheroids: shape/size effects
and benchmark results\thanksref{JQSRT}}
\author[SPB1]{N.V.~Voshchinnikov\thanksref{nvv}}
\author[SPB1]{V.B.~Il'in}
\author[Jena]{Th.~Henning}
\author[Jena]{B.~Michel}
\author[SPB2]{V.G.~Farafonov}

\address[SPB1]{Sobolev Astronomical Institute, St.~Petersburg University,
               St.~Petersburg-Peterhof, 198904 Russia}
\address[Jena]{Astrophysical Institute and University Observatory,
Fr.~Schiller University, Schillerg\"a{\ss}chen 2-3, D-07745 Jena, Germany}
\address[SPB2]{St.~Petersburg University of Aerocosmic Instrumentation,
St.~Petersburg, 190000 Russia}

\bc
({\small\it Received 10 September 1998})
\ec
\thanks[JQSRT]{Accepted for publication in
{\em Journal of Quantative Spectroscopy \& Radiat. Transfer\/}}
\thanks[nvv]{To whom all correspondence should be addressed.
Phone: (+7) 812/428 42 63;
Fax: (+7) 812/428 71 29; e-mail:nvv@aisbpu.spb.su}
\begin{abstract}
We use the separation of variables and T-matrix
methods to calculate the optical properties of homogeneous
spheroids with  refractive indices from $m = 1.3+0.0i$ up to $3+4i$.

 It is found that the extinction cross-sections for highly absorbing spheroids
are normally 1.5 -- 2 times larger than those for spheres
of the same volume.
 The albedo of the non-spherical particles rather
slightly depends on the particle shape and is mainly determined
by the imaginary part of the refractive index.
 Beginning at some size, the spheroidal particles
do not polarize the transmitted radiation independent of their shape.

We also suggest a new approach for axisymmetric particles
which would combine the strong aspects of both methods mentioned
above and
give several values of the cross-sections as  benchmarks in tabular form.
\end{abstract}
\end{frontmatter}

\section{Introduction}
In many scientific and engineering applications
prolate and oblate spheroids are appropriate  models for real particles.
 The scattering of electromagnetic
radiation by spheroids is usually calculated
by the separation of variables method \cite{ay, vf} (SVM)
and the T-matrix method \cite{bh} (TMM).
 Although both approaches are well-suited for numerical computations,
it is often difficult to give reliable error estimates.
 For non-absorbing particles, there always exists an internal check
as the absorption cross-sections
(and efficiency factors) must be equal to zero.
 Obviously, for particles consisting of absorbing materials
like silicates, graphite or metals, this test does not work.
 Despite the fact that rather elaborated methods for the calculation of
 the   optical properties of highly absorbing non-spherical particles
 can be found in literature  (see, for example, Michel et al. \cite{mich}),
these methods have never been used for detailed numerical studies.

In this paper, we consider the light scattering by homogeneous
spheroids using the most popular approaches -- SVM and TMM.
 The extinction and polarization cross-sections
for prolate and oblate spheroids of various aspect ratios $a/b$
are calculated for the refractive indices $m$ typical of
soot, graphite, iron, silicate, and water ice
and compared with the results for the particles of the same volume.
 A part of the results is presented in
 tabular form serving as benchmarks
for forthcoming calculations.
 We also describe a combined approach to calculate the light scattering
by axisymmetric particles.
 The method should allow to treat easily
(very) elongated and flattened absorbing particles.

\section{Description of methods}
\subsection{General definitions}
A spheroid (ellipsoid of revolution) is obtained by the rotation of
an ellipse around
its major axis (prolate spheroid) or its minor axis (oblate spheroid).
 The ratio of the major semiaxis $a$ to the minor semiaxis $b$
 (i.e. the aspect ratio $a/b$)
characterizes the particle shape which may vary from a nearly spherical one
($a/b \approx 1$) to a needle or a disk ($a/b \gg 1$).

We assume an incident plane wave
having the  wavelength $\lambda$. Let $\alpha$ denote the angle
between the propagation direction and the rotation axis of the spheroid 
($0^{\circ} \leq \alpha \leq 90^{\circ}$).
 
For axial propagation ($\alpha=0^{\circ}$),  there is
no polarization of transmitted radiation due to symmetry.
 If $\alpha \ne 0^{\circ}$, two  cases
of polarization of the incident radiation have to be considered:
the electric vector $\vec{E}$
is parallel (TM mode) or perpendicular (TE mode) to the plane defined
by the spheroid's rotation axis and the wave propagation vector.

The size parameter is given by
\be
x_{\rm V} = \frac{2\pi r_{\rm V}}{\lambda},
\ee
where $r_{\rm V}$ is the radius of the sphere whose volume is
equal to that of the spheroid.
 The radius $r_{\rm V}$ is defined as
\begin{equation}
r_{\rm V}^{3} = a b^{2}
             \,\,\,\,\,\,\,\,\,\,\, \mbox{for prolate  spheroids}
\end{equation}
and
\begin{equation}
r_{\rm V}^{3} = a^{2} b
             \,\,\,\,\,\,\,\,\,\,\, \mbox{for oblate  spheroids.}
\end{equation}

One usually calculates the efficiency factors $Q = C/G$
which are the ratio of the corresponding cross-sections $C$ to
the geometrical cross-section $G$ of the spheroid
(the area of the particle's shadow)
\begin{equation}
G(\alpha) = \pi b \left(a^2\sin^2\alpha
            + b^2\cos^2\alpha\right)^{1/2}
             \,\,\,\,\,\,\, \mbox{for prolate  spheroids}
\end{equation}
and
\begin{equation}
G(\alpha) = \pi a\left(a^2\cos^2\alpha
            + b^2\sin^2\alpha\right)^{1/2}
             \,\,\,\,\,\, \mbox{for oblate spheroids.}
\end{equation}

In order to compare the optical properties of the particles of different
shapes it is convenient to consider the ratios of the cross-sections
for spheroids to the geometrical cross-sections of the
equal volume spheres, $C/\pi r^{2}_{\rm V}$.
 They can be found as
\be
\frac{C}{\pi r^{2}_{\rm V}} = \frac{[(a/b)^2\sin^2\alpha
                             +\cos^2\alpha]^{1/2}}{(a/b)^{2/3}} Q
\ee
for a prolate spheroid and
\be
\frac{C}{\pi r^{2}_{\rm V}} = \frac{[(a/b)^2\cos^2\alpha
                             +\sin^2\alpha]^{1/2}}{(a/b)^{1/3}} Q
\ee
for an oblate spheroid.

The albedo of a particle can be calculated from the extinction and
scattering cross-sections
\be
\Lambda = \frac{C_{sca}}{C_{ext}}.
\ee
The dichroic polarization efficiency is defined
by the extinction cross-sections for TM and TE modes
\be
\frac{P}{\tau} = \pm \frac{C^{\rm TM}_{ext}-C^{\rm TE}_{ext}}
                          {C^{\rm TM}_{ext}+C^{\rm TE}_{ext}} \cdot 100\%,
\ee
where the upper (lower) sign is related to prolate (oblate) spheroids.
This ratio describes the efficiency of the  polarization of light
transmitted through the uniform slab consisting of non-rotating
particles of the same orientation.

The optical properties of spheroidal particles can be determined by various
methods of light scattering theory. Most frequently,
the separation of variables
method obtained by Asano \& Yamamoto \cite{ay} and Farafonov \cite{vf,v}
and the T-matrix method \cite{bh,mtm} are used.
 The calculations can be also performed with the discrete dipole
approximation (DDA) which can be applied to arbitrarily inhomogeneous
and irregular particles, but   requires large computation
time and is less accurate than the other methods (see discussion
in the papers \cite{hov, vis}).

\subsection{Asano \& Yamamoto's solution for spheroids (SVM1)}
Historically it was the first approach to the light scattering problem
for spheroids with a complex refractive index.
 The method is based on the solution to the Helmholtz equation in the
 spheroidal coordinate system.

 In their pioneering paper, Asano \& Yamamoto \cite{ay}  used
the Debye potentials to describe the electromagnetic
fields, which is similar to the Mie solution for spheres.
 The scattering coefficients then are bound
in the infinite systems of the linear
algebraic equations and can be found by solving the truncated systems.

In our comparison, we made the calculations with the numerical code
of Peter Martin (see \cite{rm, km}).

\subsection{Farafonov's solution for spheroids (SVM2)}
The principal distinction of
this solution (see \cite{vf, vf851}) from the previous one is
the special basis for the representation
of the electromagnetic fields --  a combination of
the Debye and Hertz potentials (i.e. the potentials introduced
to solve the light scattering problem for spheres and
infinitely long cylinders, respectively).

 The approach has an incontestable advantage for strongly
elongated or flattened particles.
 In this paper, the most recent version of our numerical code has been used.
It involves different methods to calculate the spheroidal functions
with the automatic choice of the most appropriate method among them.

\subsection{Solution for spheroids by the T-matrix method (TMM1 and TMM2)}
Besides SVM, another very popular approach
to solve the light scattering problem for spheroids
is the T-matrix method.
 It is mainly applied to the axisymmetric particles:
finite cylinders, spheroids, Chebyshev particles.

The main idea of the  method is the expansion of
the incident, internal and scattered radiation fields
in terms of the vector spherical harmonic functions.
 Because of the linearity of the Maxwell equations, the relations
between the expansion coefficients of
the fields are linear and given by two matrices.
Thus, solution consists of calculations of the matrix elements
being the surface integrals and inversion of a matrix.
 The advantages of the approach are simple compact codes and their high speed.
Another advantage is  also the ability to treat more the complex case
of elastic waves, the potential applicability to particles
of complex shape, and the possible analytical averaging over the particle
orientations within the expansion coefficients which
can greatly enlarge the field of the TMM applications.

We used the T-matrix codes of Barber and Hill \cite{bh}
(TMM1) and Mishchenko (TMM2; see \cite{mtm}  and references therein).

\subsection{A comparison of methods}
\begin{figure}
\resizebox{14.0cm}{!}{\includegraphics{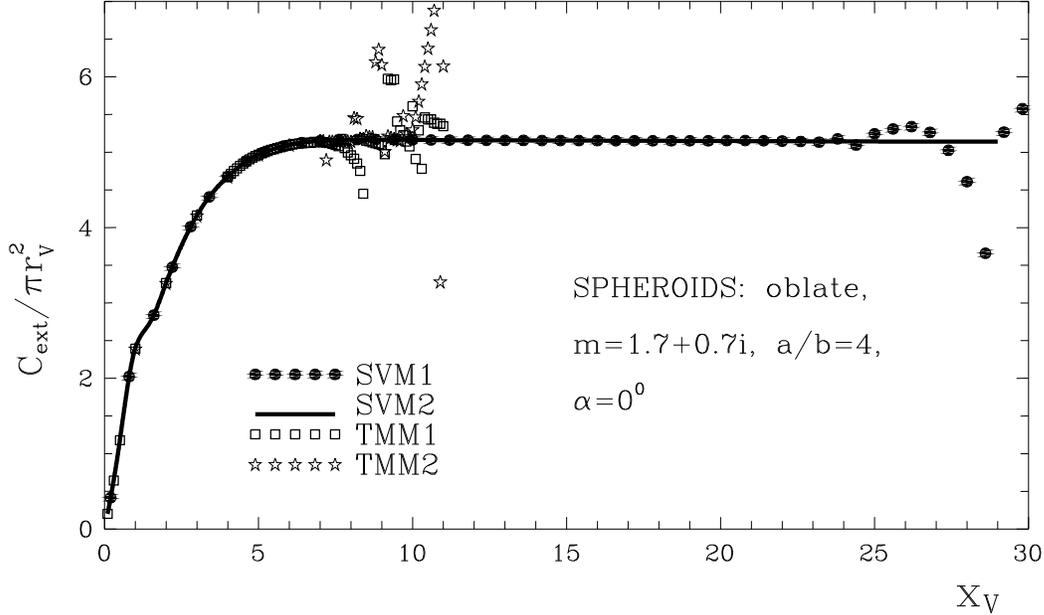}}
\caption[]{
Normalized extinction  cross-sections vs $x_V$ for
oblate spheroids with the refractive index $m=1.7+0.7i$,
$a/b =4$ and $\alpha =0^{\circ}$.
The solutions of Asano \& Yamamoto (SVM1) and Farafonov (SVM2) and the
T-matrix codes of Barber \& Hill (TMM1) and Mishchenko (TMM2)
were used in the calculations.
Note that because of the special normalization (see Eq.~(7))
the cross-sections have the limit $2 (a/b)^{2/3} \approx 5.040$
if $x_V \rightarrow \infty$.
}
\label{f1}
\end{figure}

We do not intend here to compare in detail the computational efficiency of
different methods.
The goal of this subsection is only to illustrate it on one example.
 A crude impression of the range of applicability of the
methods may be obtained from Fig.~\ref{f1} where the
normalized cross-sections are plotted for oblate spheroids
with $m = 1.7 + 0.7i$ and $a/b = 4$.
 In all four cases double precision was used.

 It should be noted that the advantage of the T-matrix codes
appears mainly for particles more spherical
than presented in the Fig.~\ref{f1}.
 On the other hand, for larger aspect ratios,
the SVM2 code becomes unrivalled.

\subsection{A new method for axisymmetric particles}
Each of the methods (SVM and TMM) has its own problems.
We suggest the following approach that could solve some of them.

The incident, scattered and internal fields are divided in two parts:
axisymmetric and non-axisymmetric ones
\be
\vec{E} = \vec{E}_{1} + \vec{E}_{2}, \ \ \ \
\vec{H} = \vec{H}_{1} + \vec{H}_{2}.
\ee
It is possible to do so that the axisymmetric part does not depend
on the azimuthal angle ($\varphi$)
and the averaging of the non-axisymmetric part
over all the azimuthal angle values gives zero \cite{vf}.
Then the scattering problem can be solved
for each of the parts separately.

The scattering problem is formulated in the integral form
like in the TMM (e.g., \cite{bh}).
 The Abraham's potentials
\be
p = E_{1\varphi} \cos\varphi, \ \ \ \
q = H_{1\varphi} \cos\varphi
\ee
are introduced for the axisymmetric part.
 A superposition of the Debye and Hertz potentials
is used for the non-axisymmetric part:
\be
\vec{E}_{2} = \vec{\nabla} \times (U \vec{e}_{z} + V \vec{r}), \ \ \ \
\vec{H}_{2} = \frac{1}{i \mu k_0} \vec{\nabla} \times \vec{\nabla}
              \times (U \vec{e}_{z} + V \vec{r})
\ee
for TE mode and
\be
\vec{E}_{2} = -\frac{1}{i \varepsilon k_0} \vec{\nabla} \times \vec{\nabla}
              \times (U \vec{e}_{z} + V \vec{r}), \ \ \ \
\vec{H}_{2} = \vec{\nabla} \times (U \vec{e}_{z} + V \vec{r})
\ee
for TM mode.
 Here $\varepsilon$ is the complex permittivity,
$\mu$ the magnetic permeability,
$k_0$ the wavevector in vacuum,
$\vec{r}$ the radius-vector,
$\vec{e}_{z}$ the unit vector,
and $i$ the complex unity.

The scalar potentials are expanded in terms of the spherical wave functions.
The coefficients of the expansions are determined from solution
to the algebraic systems similar to those obtained in the TMM (e.g., \cite{bh}).

 Thus, the new approach should combine the strong aspects of the basic methods:
the simple solution scheme typical of the TMM and
the ability to treat particles whose shape can be
very elongated or flattened as in the SVM2.
 The theoretical description of this approach is given in \cite{f}
 the development of a numerical code  realizing it  is in progress
 \cite{fih}.

\section{Numerical results}
Here we present some results illustrating
the behaviour of the optical properties of spheroidal particles
for the case of transmitted radiation.
Because this is the first consideration of
optical properties of highly absorbing non-spherical particles
beyond the Rayleigh limit, we try to demonstrate the features of such
particles in extinction, scattering and polarization.

Note that in the standard case of light scattering by spheres (Mie theory)
only the refractive index and the size parameter of particles may be varied.
 For the homogeneous spheroids, one can also study
the effects of the particle shape (the aspect ratio $a/b$) and
orientation (the angle $\alpha$) as well as that
of the polarization state of the incident
radiation (TM and TE modes) for both elongated (prolate
spheroids) and flattened (oblate spheroids) particles.

We consider  spheroids with aspect ratios $a/b = 1.1, 2, 4,$
and 10 in a fixed orientation. The first $a/b$ value illustrates
the influence of small deviations from the spherical shape.
 All the calculations presented in this Section were performed
using our new SVM2 code.

\subsection{Extinction}
\begin{figure}
\resizebox{14.0cm}{!}{\includegraphics{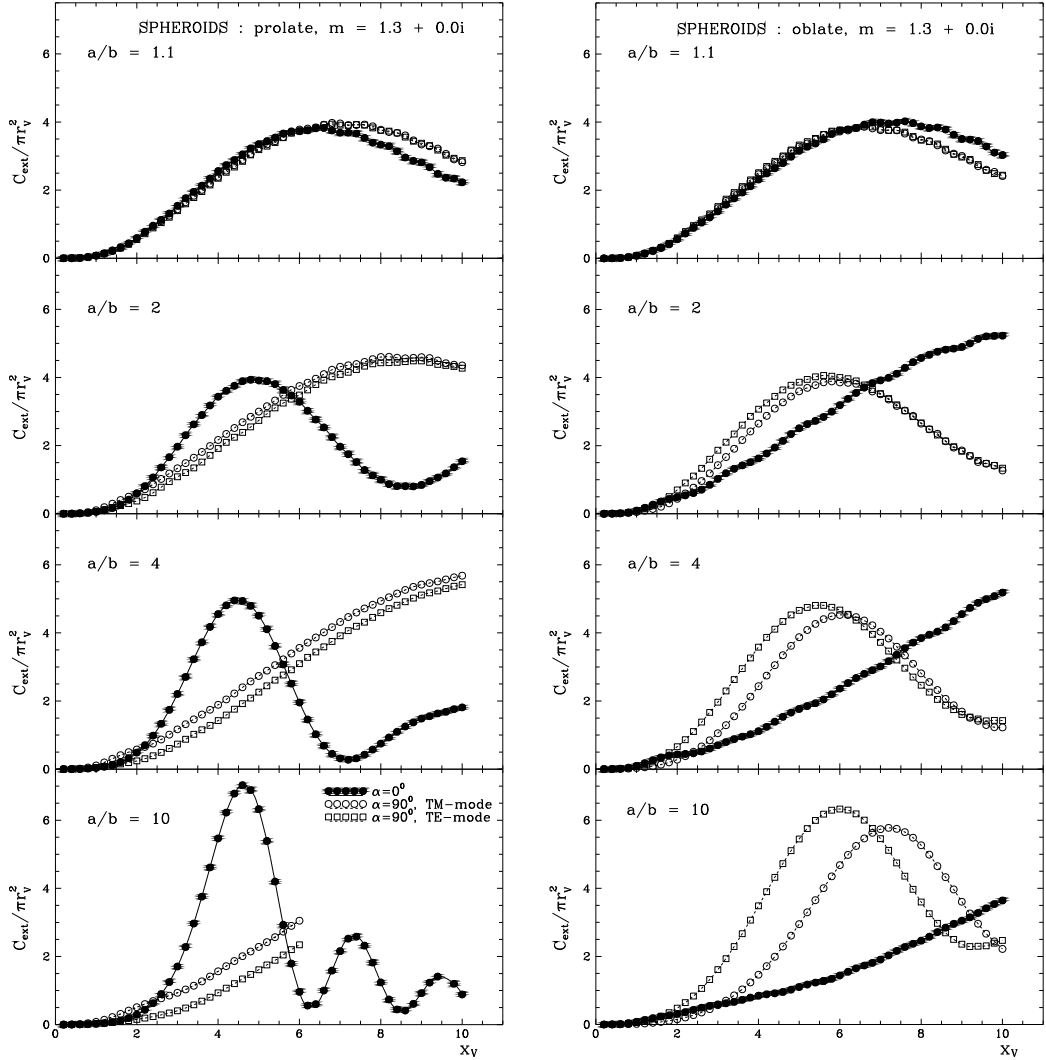}}
\caption[]{
Normalized extinction cross-sections vs $x_V$ for
prolate and oblate spheroids with $m=1.3+0.0i$.
}
\label{f2}
\end{figure}

\begin{figure}
\resizebox{14.0cm}{!}{\includegraphics{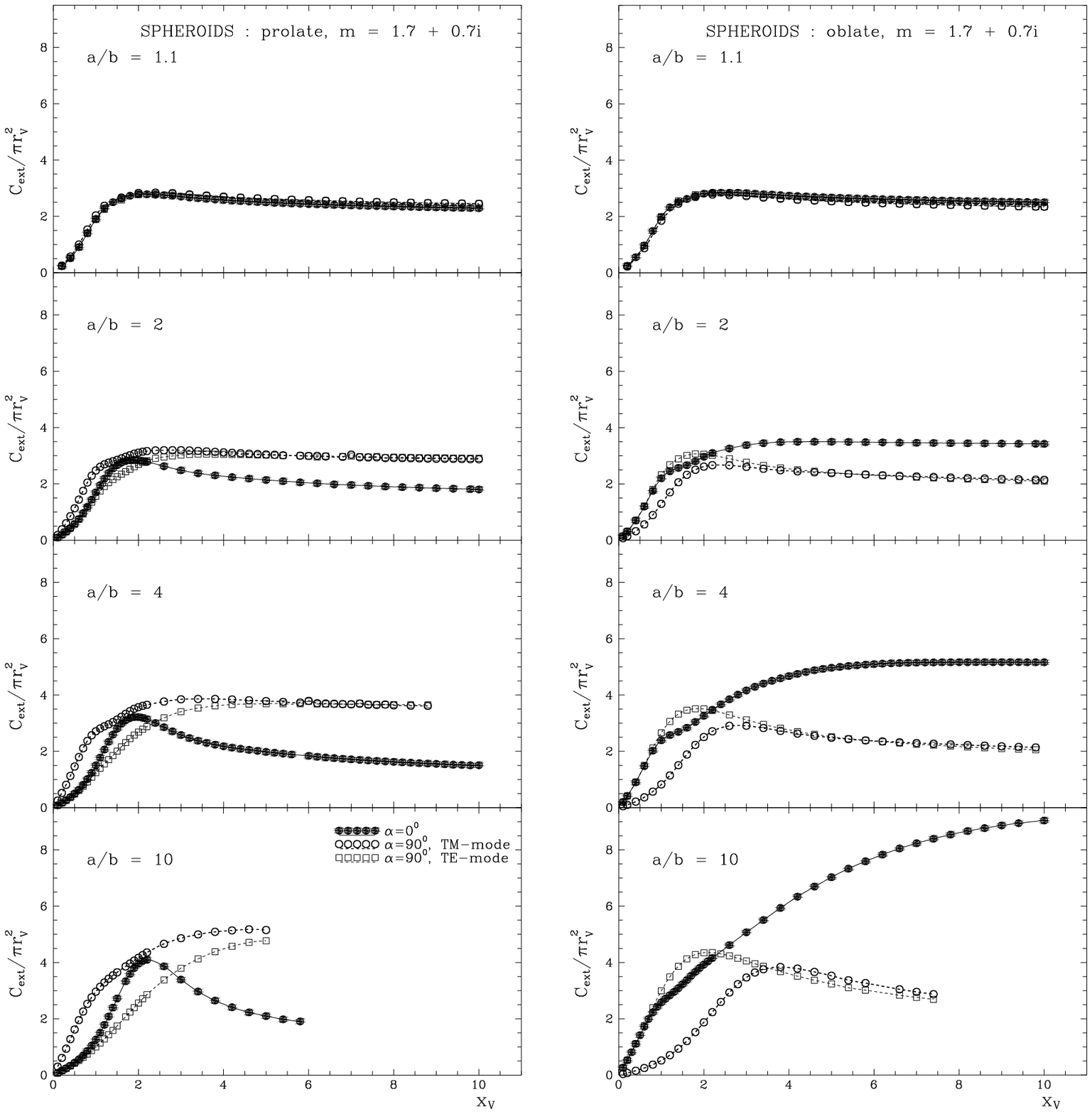}}
\caption[]{
Normalized extinction cross-sections vs $x_V$ for
prolate and oblate spheroids with $m=1.7+0.7i$.
}
\label{f3}
\end{figure}

Figures~\ref{f2} and \ref{f3} show the normalized extinction cross-sections
for non-absorbing and absorbing spheroids with different aspect ratios.
 The similar scale of the $y-$axes allows to see clearly the shape effects
for the particles of equal volume.

Note that large oscillations and the ripple-like structure
disappear for absorbing spheroids -- the curves become very smooth.
 These features depend mainly on the value of the
imaginary part of the refractive index.
 The curves $C_{\rm ext}/\pi r_{V}^{2} (x)$
for $\alpha= 0^{\circ}$ and $\alpha= 90^{\circ}$ differ more and more
when the ratio $a/b$ grows (Fig.~\ref{f3}).
 For absorbing spheroids, the position of the first (main) maximum
slightly shifts to larger values of $x_V$
with increasing $a/b$, while for non-absorbing particles,
we have the opposite case.

\begin{figure}
\resizebox{14.0cm}{!}{\includegraphics{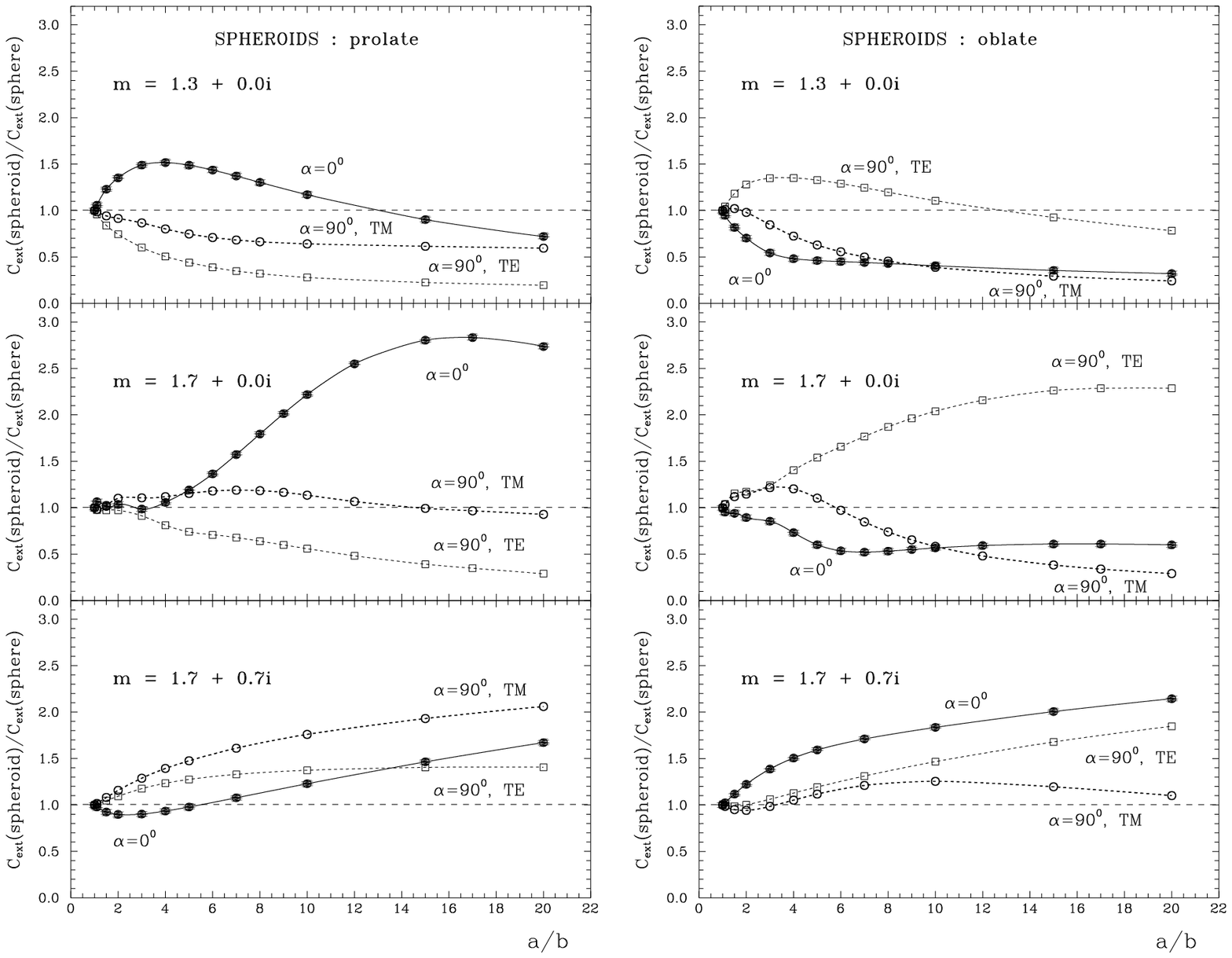}}
\caption[]{
Normalized extinction cross-sections
(relative to that for equal volume sphere) vs $a/b$
for prolate and oblate spheroids with $x_V=3$ and
$m=1.3+0.0i, 1.7+0.0i$, and $1.7+0.7i$.
}
\label{f4}
\end{figure}

 Now let us consider the particle of a fixed volume and begin
to change its shape from a sphere to a needle or disk.
 The extinction by the particle is
shown in Fig.~\ref{f4} for three values of the refractive index.
 As follows from this Figure, in almost all cases the extinction
cross-sections seem to reach their asymptotic values already
at $a/b \sim 10$.
 It is interesting that the extinction by non-spherical transparent
particles is usually smaller than that by spheres.
 The absorbing spheroids normally remove from the incident
beam 1.5 -- 2 times more energy than spheres of the same
volume.
 Though the behaviour of $C_{\rm ext}$ is refractive index/size
dependent, the conclusions we made
are rather general for absorbing spheroids.
 From Figs.~\ref{f3}, \ref{f4} one can also see that even after averaging
over particle's orientations we should get a stronger extinction
for  particles with larger aspect ratios $a/b$.

\subsection{Scattering and albedo}
If the imaginary part of the refractive index is not zero,
we can study independently the scattering and absorbing properties
of particles.
 In the case presented as an example in Fig.~\ref{f3}, the behaviour
of extinction is mainly determined by that of absorption.

\begin{figure}
\resizebox{14.0cm}{!}{\includegraphics{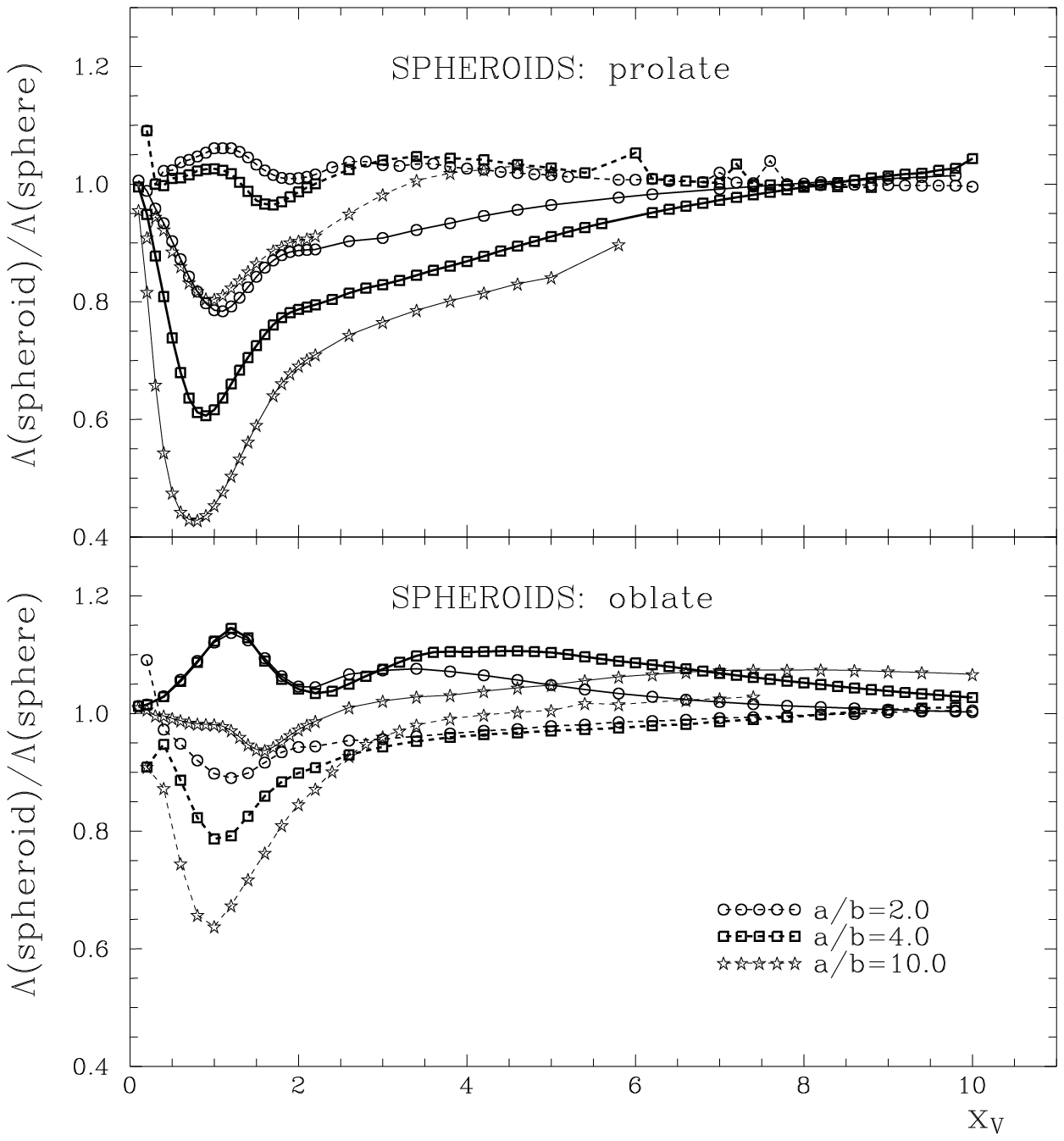}}
\caption[]{
Albedo of spheroidal particles with $m=1.7+0.7i$
normalized relative to albedo of spherical particles with
the same refractive index.
The solid lines connect the points for $\alpha = 0^{\circ}$,
the dashed lines connect the points for $\alpha = 90^{\circ}$.
}
\label{f5}
\end{figure}

The integral scattering properties of particles are characterized
by their albedo (see Eq.~(8)). This quantity depends
on the particle size and, in general, on the particle shape.
 As far as we know the shape effects have not yet been
discussed in the literature.
 Figure~\ref{f5} shows the size dependence of the albedo for spheroids with
$m=1.7+0.7i$. The calculations were made for the particles with four
aspect ratios $a/b = 1.1, 2, 4, 10$ and for two orientations
$\alpha = 0^{\circ}$ and $90^{\circ}$
(we adopt that the incident radiation is non-polarized).

It can be  clearly seen that the albedo
for non-spherical particles whose sizes are larger than
a critical value ($x_V \approx 2-3$) does not deviate more than
$\sim 20\,\%$ from that of spheres. The largest
deviations in Fig.~\ref{f5}  occur for the aspect ratio $a/b = 10$
and $\alpha = 0^{\circ}$ (prolate spheroids) and
$\alpha = 90^{\circ}$ (oblate spheroids).
 Note that for arbitrarily oriented particles (3D-alignment),
the distinction between the albedo of spheres and spheroids should be
smaller (see Fig.~11 in \cite{mtm}  where the albedo for particles
with $m=1.53+0.008i$ and $a/b=2$ is plotted).

Our calculations made for  particles with different absorption
show that the distinction of the albedo for spheres and spheroidal particles
remains rather small (within $\sim 20\,\%$) if the ratio
of the imaginary part of the refractive index to its real part
$k/n \ga 0.2 - 0.3$, which corresponds to $k \approx 0.3 - 0.4$
if $n = 1.7$.
Thus, we can conclude that the albedo of spheroidal particles
from soot, graphite, iron
very slightly depends on the particle shape.

\subsection{Polarization}
If a volume contains aligned non-spherical particles,
the initially non-polarized incident radiation
will be partially polarized
after passing the volume.
 The polarization degree can be used to estimate the alignment
degree of the particles. The simplest and at the same time extreme
case of particles' alignment is the perfect alignment of non-rotating
particles (picked fence orientation).
 The maximum polarization usually occurs when
the major axis of the particles is
perpendicular to the direction of the incident radiation
($\alpha = 90^{\circ}$).

The previous calculations of the polarization
(see \cite{vf, rm, km}) did not deal with large, highly absorbing
particles. However, such particles exist in nature and can be
aligned under terrestrial and extraterrestrial conditions.

\begin{figure}
\resizebox{14.0cm}{!}{\includegraphics{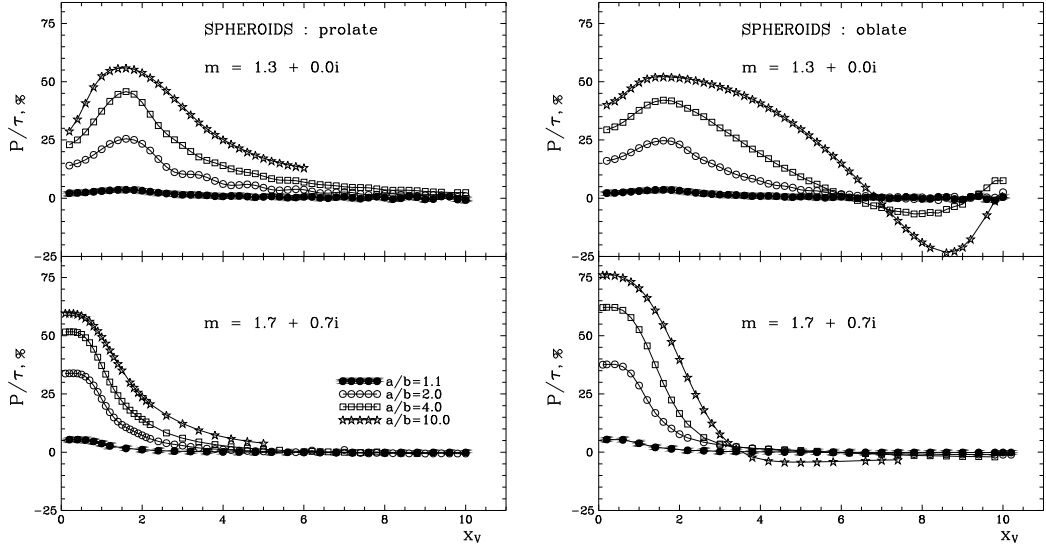}}
\caption[]{
Polarization efficiency vs $x_V$ for
prolate and oblate spheroids with $m=1.3+0.0i$ and $1.7+0.7i$,
$\alpha= 90^{\circ}$.
}
\label{f6}
\end{figure}

The behaviour of the polarization efficiency $P/\tau$ (see Eq.~(9))
for non-absorbing and absorbing spheroids is shown in Fig.~\ref{f6}
for the case when a maximum polarization is expected (picked fence
alignment, $\alpha = 90^{\circ}$).
 It is clearly seen from this Figure
that the relatively large particles produce
{\it no polarization}---independent of their shape.
 For absorbing particles, it occurs at smaller $x_V$ values
than for non-absorbing particles.
 Note also that the polarization reversal
takes place for disk-like particles.
 This effect seems to depend on the imaginary part
of the refractive index as well.

\subsection{Benchmarks}
The results of computations for spheroids
were previously published in the tabular form only in the papers
of Voshchinnikov \& Farafonov \cite{vf851,vf852,vf86}
(efficiency factors for spheroids with the refractive index $m=1.2$),
Kuik et al. \cite{khh}  and Hovenier et al. \cite{hov}
(the scattering matrices for spheroids with $m=1.5+ki, k<0.01$).
 Extensive calculations of the elements of the scattering matrices
for randomly oriented slightly absorbing spheroids with $a/b \leq 2$
were performed by Mishchenko and others (see \cite{mtm}  and
references therein).

In Tables~1~--~4, we present the normalized extinction and scattering
cross-sections for prolate and oblate spheroids with
the refractive indices $m = 2.5+0.0i, 1.7+0.7i, 2.5 +1.5i$ and $3.0 +4.0i$.
The results --- all 7 (or in a few cases 6) significant digits ---
were obtained with
{\it at least two} of four codes used (see Sects.~2.2--2.4).

\bigskip
{\bf Editor: Tables 1 - 4 are put here!!!}
\bigskip

The numbers printed by italics in Table~1
were calculated using the SVM2 code only.
 As the extinction and scattering
cross-sections were the same (with 8 and more digits),
the values presented in the Table are expected to be correct.

\section{Conclusions}
On the basis of the solutions to the light scattering problem
by the separation of variables method and the T-matrix method
we have considered the optical properties of absorbing
spheroidal particles of different aspect ratios.
 Distinct approaches allowed to get the reliable results
which may be used as benchmarks.

The following conclusions can be made from the study:

1. The extinction cross-sections for highly absorbing spheroids
are typically 1.5 -- 2 times larger than those for spheres
of the same volume.

2.  The albedo of non-spherical particles
exhibits only a weak dependence on the particle shape and is determined
mainly by the imaginary part of the refractive index.

3. For particles larger than a certain minimum size, the spheroidal particles
do not polarize the transmitted radiation independent of their shape.

\noindent{\bf Acknowledgements}

This work was supported by a grant
from the Volkswagen Foundation.

\end{document}